# A Simple Python Testbed for Federated Learning Algorithms

Miroslav Popovic
*University of Novi Sad*
*Faculty of Technical Sciences*
Novi Sad, Serbia
miroslav.popovic@rt-rk.uns.ac.rs

Marko Popovic
*RT-RK Institute for Computer Based Systems*
Novi Sad, Serbia
marko.popovic@rt-rk.com

Ivan Kastelan
*University of Novi Sad*
*Faculty of Technical Sciences*
Novi Sad, Serbia
ivan.kastelan@uns.ac.rs

Miodrag Djukic
*University of Novi Sad*
*Faculty of Technical Sciences*
Novi Sad, Serbia
miodrag.djukic@rt-rk.uns.ac.rs

Silvia Ghilezan
*University of Novi Sad & Mathematical Institute SASA*
Novi Sad, Serbia
gsilvia@uns.ac.rs

*Abstract*—Nowadays many researchers are developing various distributed and decentralized frameworks for federated learning algorithms. However, development of such a framework targeting smart Internet of Things in edge systems is still an open challenge. In this paper, we present our solution to that challenge called Python Testbed for Federated Learning Algorithms. The solution is written in pure Python, and it supports both centralized and decentralized algorithms. The usage of the presented solution is both validated and illustrated by three simple algorithm examples.

*Keywords—distributed systems, edge computing, decentralized intelligence, federated learning, Python*

I. INTRODUCTION

*Federated learning* was introduced by McMahan et al. [1] as a decentralized approach to model learning that leaves the training data distributed on the mobile devices and learns a shared model by aggregating locally computed updates. They presented FedAvg, a practical method for the federated learning of deep networks based on iterative model averaging, see Algorithm 1 FederatedAveraging in [1] on page 5. The main advantages of federated learning are: (i) it preserves local data privacy, (ii) it is robust to the unbalanced and non-independent and identically distributed (non-IID) data distributions, and (iii) it reduces required communication rounds by 10–100x as compared to synchronized stochastic gradient descent (FedSgd).

McMahan's seminal paper [1] inspired many researchers' papers and in this limited space we mention just few of them. Immediately after [1], Bonawitz et al. [2] introduced an efficient secure aggregation protocol for federated learning, and Konecny et al. [3] presented algorithms for further decreasing communication costs. More recent papers are focused on data privacy [4, 5].

TensorFlow Federated (TFF) [6], [7] is Google's framework supporting the approach introduced in [1], which provides a rich API and many examples that work well in Colab notebooks. However, TFF is a framework for applications in the cloud-edge continuum, with a heavyweight server executing in the cloud, and therefore not deployable to edge only. Besides, TFF is not supported on OS Windows, which is used by many researchers, and TFF has numerous dependencies that make its installation far from trivial.

BlueFog [8], [9] is another framework with the same limitations as TFF. In their note on page 5 in [9], BlueFog authors say that they consider deep training within high-performance data-centre clusters. Recently, Kholod et al. [10] made a comparative review and analysis of open-source federated learning frameworks for IoT, including TensorFlow Federated (TFF) from Google Inc [6], Federated AI Technology Enabler (FATE) from Webank's AI department [11], Paddle Federated Learning (PFL) from Baidu [12], PySyft from the open community OpenMined [13], and Federated Learning and Differential Privacy (FL&DP) framework from Sherpa.AI [14]. Based on the results of their analysis, they concluded that, currently, the application of these frameworks in the Internet of Things (IoTs) environment is almost impossible. In summary, at present, developing a federated learning framework targeting smart IoTs in edge systems is still an open challenge.

In this paper, we present our solution to that challenge called Python Testbed for Federated Learning Algorithms (PTB-FLA). As the word "testbed" in its title suggests, PTB-FLA was developed with the primary intention to be used as a framework for developing federated learning algorithms (FLAs), or more precisely as a runtime (or execution) environment for FLAs under development on a single computer (i.e., localhost). An important direction of our future work is to extend PTB-FLA to run on a local area network, and perhaps even to be used as a runtime in edge systems.

PTB-FLA is written in pure Python, which means that it only depends on the standard Python packages, such as the package multiprocessing, and it was intentionally written this way for the following two reasons: (i) to keep the application footprint small so to fit to IoTs, and (ii) to keep installation as simple as possible (with no external dependencies).

PTB-FLA enforces two restrictions that must be obeyed by the algorithm developers. First, a developer writes a single application program, which is later instantiated and launched by the PTB-FLA launcher as a set of independent processes whose behaviour depends on the process id. Second, a developer only writes callback functions for the client and the Funded by the European Union (TaRDIS, 101093006). Views and opinions expressed are however those of the author(s) only and do not necessarily reflect those of the European Union. Neither the European Union nor the granting authority can be held responsible for them.

server roles, which are then called by the generic federated learning algorithms hidden inside PTB-FLA.

PTB-FLA supports both centralized and decentralized federated learning algorithms. The former is as defined in [1], whereas the latter are generalized such that each process (or node) behaves as both a client and a server or more precisely it alternatively takes server and client roles from [1].

The rest of the paper is organized as follows. Section I.A presents related work. Section II presents the PTB-FLA design, Section III validates and illustrates the PTB-FLA usage by three simple algorithm examples, and Section IV concludes the paper.

*A. Short Discussion of Closely Related Work*

The word "testbed" in the name PTB-FLA may be misleading, it was selected by ML & AI developers in our project, because they see PTB-FLA as an "algorithmic" testbed where they can plugin and test their FLAs. However, PTB-FLA is a federated learning framework and not a system testbed, such as the one that was used for testing the system based on PySyft in [15].

Another important point that needs clarification is that PTB-FLA is just a FL framework, and it is not a complete system such as CoLearn [16] and FedIoT [17]. CoLearn is an FL system based on the open-source Manufacturer Usage Description (MUD) implementation osMUD and the FL framework PySyft, whereas FedIoT is a system for realistic IoT devices (e.g., Raspberry PI) that comprises a specialized FL framework for IoT cybersecurity named FedDetect.

PTB-FLA is an early work in progress. At this time, it only executes on a localhost, still we can compare its design principles with PySyft and FedDetect. PySyft is centralized whereas PTB-FLA supports both centralized and decentralized FLAs. FedDetect is both centralized and specialized whereas PTB-FLA is generic.

We can also compare the target edge systems. Currently, both CoLearn and FedIoT are edge system comprising computers and laboratory IoT devices like Raspberry PI, whereas PTB-FLA on its roadmap also has swarms of just IoT devices (without computers) that may use MicroPython as an OS, which is becoming common in embedded systems (see an interesting toy example in [18]). Perhaps the most challenging edge system on PTB-FLA roadmap is a swarm of LEO satellites, which is one of the use cases in TaRDIS project [19].

In summary, the problem that we are attempting to solve is how to construct a FL framework that is well-structured, generic, and based on restricted programming (and therefore easy to formally verify – this is also one of the stops on the PTB-FLA roadmap), and we think this is important because it leads to an exciting roadmap we briefly sketched. PTB-FLA is just a first step, and we hope it's in the right direction.

## II. PTB-FLA DESIGN

This section presents the PTB-FLA design details. For brevity, the term *system based on PTB-FLA* is abbreviated as the term *PTB-FLA system*. The next subsections present the PTB-FLA system architecture (Subsection II.A), the PTB-FLA API (Subsection II.B), and the PTB-FLA system operation (Subsection II.C).

*A. PTB-FLA System Architecture*

The PTB-FLA system architecture, see Fig. 1, consists of the application launcher process $s$, the distributed application $A = \{a_1, a_2, …, a_n\}$, which is a set of application program instances $a_i$, and the distributed testbed $T = \{t_1, t_2, …, t_n\}$, which is a set of testbed instances $t_i$, where $i = 1, 2, …, n$, and $n$ is the number of instances in both $A$ and $T$.

The system starts as follows. Once the launcher process $s$ is manually started from the command line interface (CLI), it instantiates $n$ application program instances $a_i$, $i = 1, 2, …, n$, and launches them as $n$ independent processes (in Fig. 1 this is illustrated as a set of rays radiating from $s$). Each application program instance $a_i$ in turn creates its testbed instance $t_i$. At the end, the testbed instances conduct the startup handshake by exchanging hello messages (details in II.C).

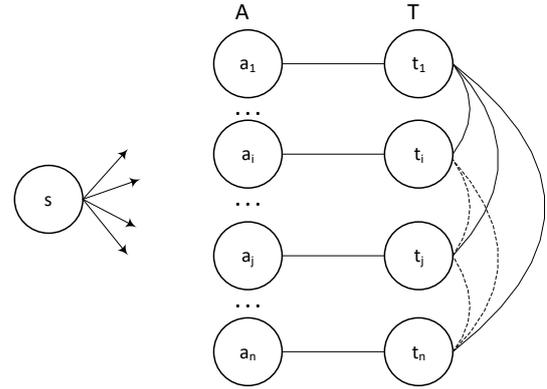

Legend: s – application launcher, A – application, ai – application program instance, T – testbed, ti – testbed program instance.

Fig. 1. Block diagram of the PTB-FLA system architecture.

During normal system operation, the distributed application $A$ uses the distributed testbed $T$ to execute the distributed algorithm, which is specified by the callback functions within the application program (i.e., in the application Python modules). PTB-FLA supports both centralized and decentralized federated learning algorithms by providing the API functions that implement the generic centralized algorithm and the generic decentralized algorithm, named fl_centralized and fl_decentralized, respectively.

The distributed federated learning algorithm (either centralized or decentralized) is executed as follows. Each instance $a_i$ prepares its input data for the generic API function based on its command line arguments (including its identification $i$, the number of instances $n$, etc.) and then calls the desired generic API function (either fl_centralized or fl_decentralized) on its testbed instance $t_i$.

The testbed instance $t_i$ in turn plays its role (determined by its id $i$) in the generic algorithm by exchanging messages with other testbed instances and by calling the associated callback function at the right point of the generic algorithm (details in II.C). In case of a centralized algorithm, the graph of testbed instances takes the form of a star, whereas in the case of a decentralized algorithm it takes the form of clique (or complete graph). In Fig. 1, on the right side, the solid oval edges connecting $t_1$ (a server) with other testbed instances (clients) illustrates the former case, whereas all the branches (solid and dashed) illustrate the latter case.

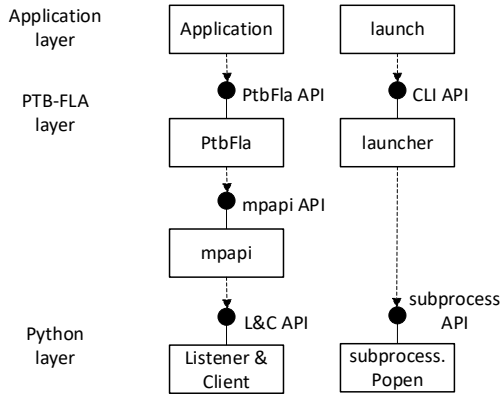

Fig. 2. UML class diagram of the PTB-FLA system architecture.

The PTB-FLA system architecture comprises three layers: the distributed application layer on top (comprising the application modules and the console script launch), the PTB-FLA layer (comprising the class PtbFla in the module ptbfla and the modules mpapi and launcher), and the Python layer (including classes Process, Queue, and Listener & Client from the package multiprocessing and Popen from the package subprocess).

As shown in Fig. 2, the console script launch uses the module launcher (which in turn uses Popen) to launch the distributed application comprising $n$ independent processes, $p_i$, $i = 1, 2, …, n$, where each $p_i$ comprises the corresponding pair of instances $(a_i, t_i)$ and executes in a separate terminal (i.e., window). On the other hand, the application module uses the PtbFla API (comprising PtbFla functions) to create or destroy a testbed instance (by calling the constructor or the destructor) and to conduct its role in the distributed algorithm execution (by calling the API function fl_centralized or the API function fl_decentralized).

The API functions fl_centralized and fl_decentralized, within an instance $t_i$, use the module mpapi (mpapi is the abbreviation of the term *message passing API*) to communicate with other instances. The module mpapi in turn instantiates the Python multiprocessing classes Listener and Client to create the mpapi server and the mpapi client, which are hidden with the module mpapi and should not be confused with the server and client roles in the federated learning algorithms.

The mpapi API is strictly an internal API providing services to PtbFla only, and it should never be used by the distributed algorithms' developers, instead they should only use the PtbFla API in their application program modules.

*B. PtbFla API*

The PtbFla API comprises the following four functions (the variable after "/" is the function return value):

1. PtbFla(*noNodes*, *nodeId*, *flSrvId*=0) / None
2. fl_centralized(*sfun*, *cfun*, *ldata*, *pdata*, *noIters*=1) / *ret*
3. fl_decentralized(*sfun*, *cfun*, *ldata*, *pdata*, *noIters*=1) / *ret*
4. PtbFla() / None

The first is the constructor that is called as a global function and does not have a return value, the second and the third are member functions that are called on the instance of PtbFla, and the fourth is the destructor that is called implicitly by the garbage collector or explicitly when deleting an object.

The arguments are as follows: *noNodes* is the number of nodes (or processes), *nodeId* is the node identification, *flSrvId* is the server id (default is 0; this argument is used by the function fl_centralized), *sfun* is the server callback function, *cfun* is the client callback function, *ldata* is the initial local data, *pdata* is the private data, and *noIters* is the number of iterations that is by default equal to 1 (for the so called one-shot algorithms), i.e., if the calling function does not specify it, it will be internally set to 1. The return value *ret* is the node final local data. Data (*ldata* and *pdata*) is application specific.

Typically, local data (*ldata*) is a machine learning model, whereas the private data (*pdata*) is a training data that is used to train the model. For example, in case of a simple linear regression i.e., straight-line fit to data, $y = ax + b$, the machine learning model is the pair of coefficients $(a, b)$ where $a$ is the slope and $b$ is the intercept, whereas the training data is the given array of points i.e., pairs $(x_i, y_i)$, $i = 1, …, n$, where $n$ is the number of points.

Normally, the testbed instances only exchange the local data (i.e., their local machine learning models) and they never send out the private data (that is how they guarantee the training data privacy). The private data is only passed to callback functions (within the same process instance) to immediately set them in their working context.

Note that PTB-FLA at this time has a simple startup that does not separate instances private data, but the startup in the future distributed PTB-FLA version will do that.

*C. PTB-FLA Operation*

This subsection provides an overview of the PTB-FLA operation by presenting the following three most important scenarios: (i) the system startup handshake, (ii) the generic centralized one-shot FLA (federated learning algorithm) execution, and (iii) the generic decentralized one-shot FLA execution.

The system startup handshake has two phases, see Fig. 3. In the first phase, the instance $a_1$ is waiting to receive $(n - 1)$ Hello messages from all other instances $a_i$, $i = 2, …, n$, and in the second phase, the instance $a_1$ broadcasts the message Hello to all other instances (note: conceptually, the index $i$ takes values from 1 to $n$, whereas in PTB-FLA Python implementation it goes from 0 to $n - 1$).

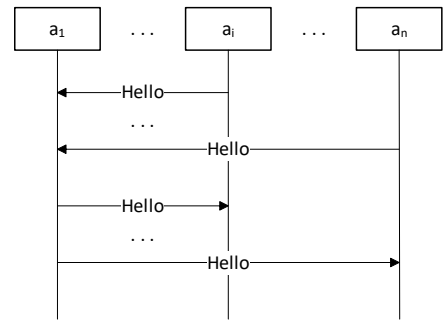

Fig. 3. The system startup handshake.

The generic centralized one-shot FLA has three phases, see Fig. 4. Let's assume that the instance $a_1$ is the server and the other instances $a_i$, $i = 2, …, n$, are the clients. In the first phase, the server broadcasts its local data to the clients, which in their turn call their callback function to get the update data and store the update data locally.

In the second phase, the server receives the update data from all the clients, and in the third phase, the server calls its callback function to get its update data (e.g., aggregated data) and stores it locally. Finally, all the instances return their new local data as their results.

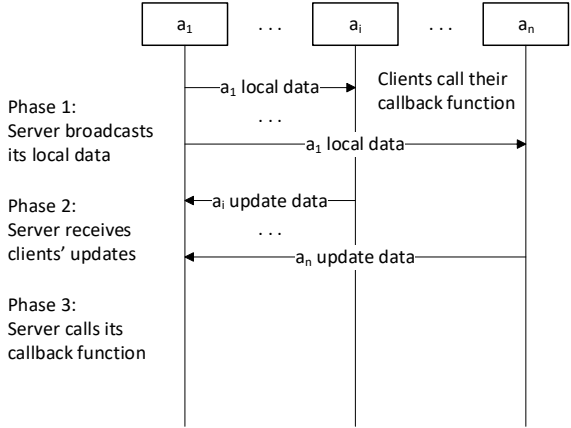

Fig. 4. The generic centralized one-shot FLA execution.

Unlike the generic centralized FLA that uses the single field messages carrying data (local or update), the generic decentralized FLA, being more complicated, uses the three field messages carrying: the messages sequence number (corresponding to the algorithm's phase number), the message source address (i.e., the source instance network address), and the data (local or update).

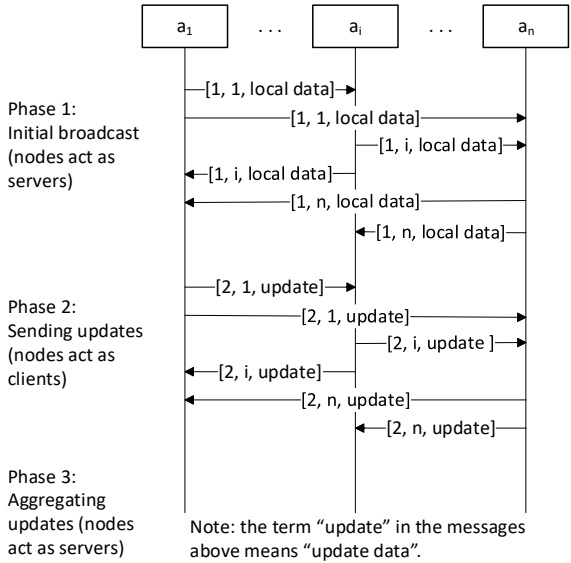

Fig. 5. The generic decentralized one-shot FLA execution.

The generic decentralized one-shot FLA has three phases, see Fig. 5. In the first phase, see the top of Fig. 5, each instance acts as a server, and it sends its local data to all its neighbours. These messages have the sequence number 1, and each instance sends $(n - 1)$ such messages. Note that each instance is also the destination for $(n - 1)$ such messages.

In the second phase, see middle of Fig. 5, each instance acts as a client, and it may receive either a message with the sequence numbers 1 (sent in the first phase) or 2 (sent during the second phase). If the instance receives a message from the second phase, it just stores it in a buffer for later processing, whereas if the instance receives a message from the first phase, it calls the client callback function to get the update data, and then sends the reply to the message source. In the reply, the instance sets the message fields as follows: the field sequence number to 2, the field message source address to its own address, and the field data to update data. Note that during the second phase, the instance does not update its local data, it just passes the update data it got form the client callback function.

Since messages are sent asynchronously, they may be received in any order. Note that for the simplicity of presentation, Fig. 5 shows a scenario where all the instances receive the messages in the phase order. However, if an instance receives the messages out of the phase order, it uses the buffer to process them in the phase order.

The second phase is completed after the instance received and processed all $2(n - 1)$ messages (from both phases). In the third phase, each instance again acts as a server, and it calls the server callback function to get its update data (e.g., aggregated data) and stores it locally. Finally, all the instances return their new local data as their results.

### III. PTB-FLA VALIDATION

This section validates and illustrates PTB-FLA usage by three simple algorithm examples (see III.A, III.B, and III.C).

#### A. Example 1: Federated Map

This example is analogous to the McMahan's federated learning example for averaging the number of sensors readings above the given threshold, see pp. 50-51 in [7].

| Algorithm 1. The algorithm example 1 |
|---|
| 01: example1(*noNodes*, *nodeId*, *flSrvId*) |
| 02: // Create PtbFla object |
| 03: *ptb* = PtbFla(*noNodes*, *nodeId*, *flSrvId*) |
| 04: // Set localData for FL server/clients as follows |
| 05: if *nodeId* == *flSrvId* then |
| 06: *localData* = 69.5 // Set the threshold |
| 07: else   // Set the client readings |
| 08: *localData* = 68.0 |
| 09: if *nodeId* == *noNodes* – 1 then |
| 10: *localData* = 70.5 |
| 11: // Call fl_centralized with *noIterations* = 1 (default) |
| 12: *ret* = *ptb*.fl_centralized(servercb, clientcb, *localData*, None) |
| 13: clientcb(*localData*, *privateData*, *msg*) |
| 14: *clientReading* = *localData* |
| 15: *threshold* = *msg* |
| 16: *tmp* = 0.0 |
| 17: if *clientReading* > *threshold* then |
| 18: *tmp* = 1.0 |
| 19: return *tmp* |
| 20: servercb(*privateData*, *msgs*) |
| 21: *listOfIsOverAsFloat* = *msgs* |
| 22: return sum(*listOfIsOverAsFloat*) / len(*listOfIsOverAsFloat*) |

The main function example1: (i) creates the object *ptb* as an instance of the class PtbFla, (ii) sets the initial local data of instance according to its *nodeId*, and (iii) calls the API function fl_centralized on *ptb*. The initial local data for the server is the given threshold (69.5), whereas the initial local data for the clients are their sensor readings, which have the value 68.0 (below the threshold) for all the clients except the last one whose reading is 70.5 (above the threshold).

The client callback function clientcb: (i) receives the client local data (its sensor reading) through the argument *localData* and the server's local data (the threshold) through the argument *msg* that is the message the client received from the server, (ii) sets the variable *tmp* to 0.0 if the reading is below the threshold or to 1.0 otherwise, and (iii) returns *tmp* to the generic function fl_centralized, which in turn forwards the *tmp* to the server. The server in turn collects all the client replies into a list and passes this list to the server callback function.

The server callback function servercb receives this list through the argument *msgs*, and in turn returns the fraction of sensor readings that are above the threshold.

*B. Example 2: Centralized Data Averaging*

This example is analogous to the McMahan's federated learning example for averaging the client models, see pp. 19-27 in [7].

| Algorithm 2. The algorithm example 2 |
|---|
| 01: example2(*noNodes*, *nodeId*, *flSrvId*) |
| 02: // Create PtbFla object |
| 03: *ptb* = PtbFla(*noNodes*, *nodeId*, *flSrvId*) |
| 04: // Set localData for FL server/clients as follows |
| 05: *localData* = [*nodeId*+1] |
| 06: // Call fl_centralized with noIterations = 10 |
| 07: *ret* = *ptb*.fl_centralized(servercb, clientcb, *localData*, None, 10) |
| 08: clientcb(*localData*, *privateData*, *msg*) |
| 09: return [(*localData*[0] + *msg*[0])/2] |
| 10: servercb(*privateData*, *msgs*) |
| 11: *tmp* = 0.0 |
| 12: for *lst* in *msgs*: |
| 13: *tmp* = *tmp* + *lst*[0] |
| 14: *tmp* = *tmp* / len(*msgs*) |
| 15: return [*tmp*] |

Like in the previous example, the main function example2 creates the object *ptb*, sets the initial local data of an instance, and calls the function fl_centralized on *ptb*. The initial local data in this example is a simple model that is encoded as a list with a single element that characterizes client behaviour (e.g., an average value of some variable). Of course, the model at the server is more authoritative than models at the clients.

The client callback function clientcb averages this client local model and the server's model received through the argument *msg* i.e., it returns the list whose element is the average of the elements from this client local list and the list in *msg*.

The server callback function servercb averages all the client models, which it receives through the argument *msgs* i.e., it returns the list whose element is the average of the elements of all the lists in *msgs*.

As expected, the local data models, i.e., the elements in the lists are converging through the iterations to an average value. Here we define the point in which the elements converged as the iteration in which the difference between an element and the average value is less than 0.02, for all elements.

Fig. 6 shows the convergence of the local data models for this example. The point in which the elements converged is the iteration 5, and the average value is 1.75. The model in the first instance converged in the first iteration, whereas the models in the second and third instance asymptotically approach the average from below and above, respectively.

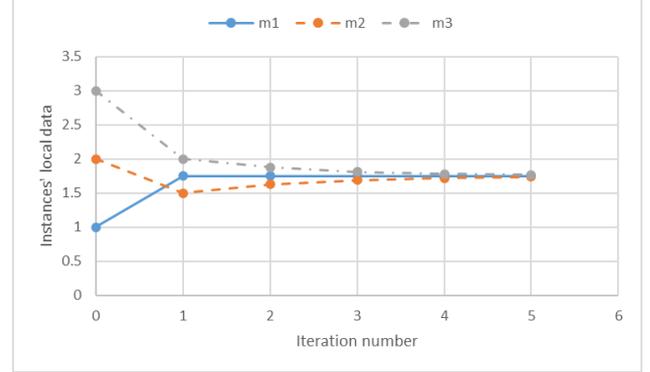

Fig. 6. Local data convergence for the centralized FLA in the example 2.

The average value 1.75 is not the simple average of the initial values of instances 1, 2, and 3 (which is 2), but is as expected because the model at the first instance (the server) is more authoritative than models at the other two (the clients) i.e., it has greater influence. Therefore, the average value 1.75 is somewhat closer to the server's initial value 1 than the simple (unweighted) average value 2.

*C. Example 3: Decentralized Data Averaging*

The pseudo code for this example, see Algorithm 3, is practically identical as for the example 2.

| Algorithm 3. The algorithm example 3 |
|---|
| 01: example3(*noNodes*, *nodeId*) |
| 02: // Create PtbFla object |
| 03: *ptb* = PtbFla(*noNodes*, *nodeId*) |
| 04: // Set localData for FL server/clients as follows |
| 05: *localData* = [*nodeId*+1] |
| 06: // Call fl_decentralized with noIterations = 10 |
| 07: *ret* = *ptb*.fl_decentralized(servercb, clientcb, *localData*, None, 10) |
| 08: clientcb(*localData*, *privateData*, *msg*) |
| 09: return [(*localData*[0] + *msg*[0])/2] |
| 10: servercb(*privateData*, *msgs*) |
| 11: *tmp* = 0.0 |
| 12: for *lst* in *msgs*: |
| 13: *tmp* = *tmp* + *lst*[0] |
| 14: *tmp* = *tmp* / len(*msgs*) |
| 15: return [*tmp*] |

The main difference in the pseudo code for this example is in the line 7, where the function fl_decentralized is called instead of the function fl_centralized. The other difference is that the variable *flSrvId* is not used, therefore the lines 1 and 3 are different. Note that callback functions (lines 8-15) are identical, but here they are called from the API function fl_decentralized, so the overall behaviour is of course different.

Fig. 7 shows the convergence of the local data models for this example. The point in which the elements converged is the iteration 3, and the average value is 2.0. The model in the second instance converged in the first iteration, whereas the models in the first and third instance asymptotically approach the average from below and above, respectively.

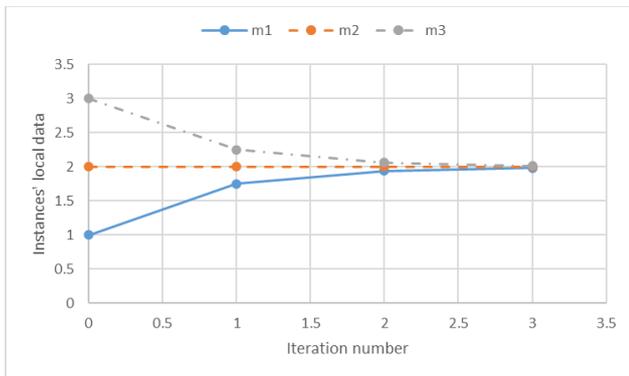

Fig. 7. Local data convergence for the decentralized FLA in the example 3.

The average value 2.0 is equal to the simple average of the initial values of instances 1, 2, and 3 (which is 2), and this is as expected because all models have equal authority i.e., they have equal influence. Therefore, the resulting average value 2.0 is equal to the simple (unweighted) average value 2.

When comparing the points where the elements converged, we see that the decentralized algorithm in the example 3 required significantly less iterations to converge than the centralized algorithm in the example 2 (we say significantly because 3 iterations is almost half of 5 iterations).

## IV. Conclusion

In this paper, we developed the federated learning framework targeting smart IoTs in edge systems called Python Testbed for Federated Learning Algorithms (PTB-FLA), with the primary intention to be used as a framework for developing FLAs on a single computer. The solution is written in pure Python, and it supports both centralized and decentralized algorithms. The PTB-FLA usage is both validated and illustrated by three simple algorithm examples.

The main PTB-FLA advantages are the following: (i) it keeps the application footprint small so to fit to smart IoTs and (ii) it keeps the installation as simple as possible (with no external dependencies).

The shortcomings of the current PTB-FLA are: (i) the maximum number of nodes and edges is limited by the available localhost hardware and OS resources, (ii) the networking aspect is not supported, and (iii) it's not validated in any way to work on real IoT devices.

In our future work we plan: (i) to develop some commonly used FLAs using PTB-FLA essentially by repacking their sequential code into the client and server callback functions, and (ii) to extend PTB-FLA to run on a local area network, and perhaps even to be used as a runtime in edge systems.

## Acknowledgment

We are grateful to Dragana Bajovic, Claudia Soares, and Panagiotis Trakadas for their valuable presentations and discussions on federated learning algorithms, as well as to all the colleagues in TaRDIS project for the friendly atmosphere and great collaboration.